\documentclass[12pt]{article}
\usepackage{epsfig}
\usepackage{amsfonts}

\newcommand{\Li}{\mbox{Li}}

\begin{document}

\begin{center}
{\Large \bf Pion form factor and QCD sum rules: \\ case of axial current.}
\\ \vspace*{5mm} V.V.Braguta$^{a}$ and A.I.Onishchenko$^{b}$
\end{center}

\begin{center}
a) Institute for High Energy Physics, Protvino, Russia \\
\vspace*{0.5cm} 
b) Department of Physics and Astronomy \\ Wayne State University,
Detroit, MI 48201, USA \\
\end{center}

\vspace*{0.5cm}

\abstract{
We present an analysis of QCD sum rules for pion form factor in next-to-leading
order of perturbation theory for the case of axial-vector pion currents.
The theoretical predictions for $Q^2$-dependence of pion form factor are in good agreement
with available experimental data. It is shown, that NLO corrections are large in
this case and should be taken into account in rigorous theoretical analysis. 
}\\

\section{Introduction}

The study of electromagnetic form factors of hadrons already has a long history. The first
applications appeared right after the realization, that perturbative QCD may be also applied
in studies of exclusive processes with high momentum transfer  \cite{Lepage:1980,Efremov:1979,Chernyak:fk}. 
However, later comparison of pQCD predictions with data lead to conclusion, that at moderate momentum transfers 
(typically of the  order of few GeV) the power corrections, otherwise known as soft or end-point
contributions should come into play\footnote{See, for example, discussion in \cite{hardsoft}.}.   

In this paper we are using the framework of QCD sum rules \cite{QCDSR} to consistently account for
both hard scattering and soft wave function overlap contributions to pion electromagnetic form factor. 
Within this approach the soft contribution is dual to the lowest-order triangle diagram, while the hard contribution
is given by diagrams having higher order in coupling constant $\alpha_s$ and as a consequence suppressed
relative to soft contribution with additional factor $\alpha_s/\pi\sim 0.1$. This extra suppression 
is in a complete agreement with asymptotic behavior of the pion electromagnetic form factor, calculated
in pQCD \cite{Lepage:1980,Efremov:1979,Chernyak:fk}:
\begin{eqnarray}
F_{\pi}^{\mathbf{hard}}(Q^2) = \frac{8\pi\alpha_s (Q^2)}{9}\int_0^1 dx \int_0^1 dy
\frac{\phi_{\pi}(x)\phi_{\pi}(y)}{xyQ^2} = \frac{8\pi\alpha_s f_{\pi}^2}{Q^2},
\label{pQCDLO}
\end{eqnarray}
where the last equality holds for the asymptotic pion distribution amplitude.
At asymptotically high $Q^2$ the ${\cal O}(\alpha_s/\pi)$ suppression of hard contribution is
more than compensated by its slower decrease with $Q^2$. However, such  compensation does not
occur in the region of moderate momentum transfer, where the soft contribution, scaling as $1/Q^4$ becomes 
large and can compete in strength with hard contribution. 

The pion electromagnetic form factor was studied using many different frameworks, like
QCD sum rules \cite{Ioffe:ia,Ioffe:qb,Nesterenko:1982gc,Nesterenko:1984tk}, light-cone sum rules
\cite{Braun:ij,Braun:1999uj,Bijnens:2002mg}, sum rules with nonlocal condensates \cite{Bakulev:ps} and NLO pQCD approach 
\cite{Field:wx,Dittes:aw,Khalmuradov:1984ij,Braaten:yy,Melic:1998qr}, 
improved by inclusion of finite size corrections to
pion distribution function \cite{Botts:kf,Li:1992nu,Jakob:1993iw}. There are also estimates
of pion electromagnetic form factor based on use of pseudoscalar pion interpolating currents
\cite{Forkel:1994pf,Faccioli:2002jd,Braguta:2003hd}. 

In what follows we will consider NLO three-point QCD sum rules for pion form factor, where axial currents 
are used as pion interpolating currents. The main result of this paper is the explicit analytical expression 
for QCD radiative corrections to double spectral density, entering QCD sum rule predictions for pion 
electromagnetic form factor. It should be noted, that up to the moment there are only radiative corrections 
computed for reduced spectral density 
\cite{Bakulev1,Bakulev2}.

The obtained results for pion electromagnetic form factor calculation are
in good agreement with existing experimental data. Here, we would like to stress that an 
inclusion of radiative corrections is very important, as only in such a way we can simultaneously account 
for both hard and soft contributions. Moreover, these corrections are large numerically and thus
should be accounted for in theoretical predictions.

The paper is organized as follows. In section 2 we describe our framework and give explicit
expressions for next-to-leading order corrections to double spectral density. Section 3
contains our numerical analysis. Finally, in section 4 we draw our conclusions.

\section{Derivation of QCD sum rules}   

To determine pion electromagnetic form factor we will use the method of three-point QCD sum rules.
Here, to describe charged pion state we choose an action of axial interpolating current on a vacuum
state. The vacuum to pion transition matrix element of axial current is 
defined by
\begin{eqnarray}
\langle 0|\bar u\gamma_5  \gamma_{\mu} d|\pi^{-} (p) \rangle = 
i f_{\pi} p_{\mu},
\end{eqnarray}
where $f_{\pi} = 131$ MeV. 
Next, the pion electromagnetic form factor, we are planning to determine, is given by hadronic matrix element
of electromagnetic current:
\begin{eqnarray}
\langle\pi (p')|j_{\mu}^{\mathbf{el}}|\pi (p)\rangle = 
F_{\pi}(Q^2)(p_{\mu} + p'_{\mu}),
\end{eqnarray}
where $j_{\mu}^{\mathbf{el}} = e_u\bar u\gamma_{\mu}u + e_d\bar d\gamma_{\mu}d$,
the momenta of initial and final state pions were denoted by $p$, $p'$ and $Q^2 = -q^2$ ($q = p -p'$)
is square of momentum transfer. 

Within the framework of QCD sum rules an expression for pion electromagnetic form factor
follows from an analysis of corresponding three-point correlation function:
\begin{eqnarray}
\Pi_{\mu \alpha \beta} (p,p',q) = i^2\int dx dy e^{i (p'\cdot x - p\cdot y)}
\langle 0|T\{\bar u(x)\gamma_5 \gamma_{\alpha} d(x), j_{\mu}^{\mathbf{el}}(0), 
(\bar u(y)\gamma_5 \gamma_{\beta} d(y) )^+  |0\rangle
\label{correlator}
\end{eqnarray}

\begin{figure}[ht]
\begin{center}
\includegraphics[scale=0.5]{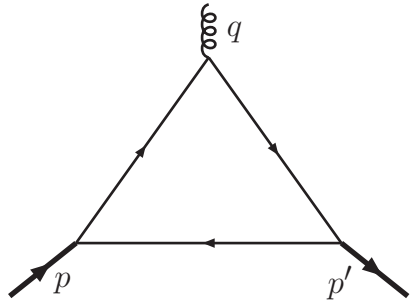} 
\caption{LO diagram}
\label{LOdiagram}
\end{center}
\end{figure}

This correlation function contains a lot of different tensor structures.
The scalar amplitudes $\Pi_i$ in front of different Lorentz structures
are the functions of kinematical invariants, i.e. $\Pi_i = \Pi_i (p^2,p'^2,q^2)$.
The calculation of QCD expression for three-point correlator is done through the use of operator product
expansion (OPE) for the T-ordered product of currents. As a result of OPE one obtains besides leading perturbative
contribution also power corrections, given by vacuum QCD condensates. We will return to the discussion of QCD
expression for three-point correlation function in a moment. Now let us discuss the physical part of our 
sum rules.  The connection to hadrons in the framework of QCD sum rules is obtained by matching the resulting
QCD expressions for current correlators with spectral representation, following the structure of
double dispersion relation at $q^2\leq 0$:
\begin{eqnarray}
\Pi_{\mu\alpha\beta}(p_1^2, p_2^2, q^2) = \frac{1}{(2\pi)^2}
\int\frac{\rho_{\mu\alpha\beta}^{\rm phys}(s_1, s_2, Q^2)}{(s_1 - p_1^2)(s_2 - p_2^2)}ds_1 ds_2
+ \mbox{subtractions}. \label{disp_phys}
\end{eqnarray} 
Assuming that the dispersion relation (\ref{disp_phys}) is well convergent, the physical
spectral functions are generally saturated by the lowest lying hadronic states plus
a continuum starting at some thresholds $s_1^{th}$ and $s_2^{th}$:
\begin{eqnarray}
\rho_{\mu\alpha\beta}^{\rm phys}(s_1, s_2, Q^2) &=& \rho_{\mu\alpha\beta}^{\rm res}(s_1, s_2, Q^2) + \nonumber \\
&& \theta (s_1 - s_1^{th})\cdot\theta (s_2 - s_2^{th})\cdot\rho_{\mu\alpha\beta}^{\rm cont}(s_1, s_2, Q^2),
\end{eqnarray}
where
\begin{eqnarray}
\rho_{\mu\alpha\beta}^{\rm res}(s_1, s_2, Q^2) &=& 
\langle 0|\bar u\gamma_5\gamma_{\alpha} d|\pi^{-}(p')\rangle 
\langle\pi^{-}(p')|j_{\mu}^{\mathbf{el}}|\pi^{-}(p)\rangle 
\langle\pi^{-}(p)|(\bar u\gamma_5\gamma_{\beta} d)^{+}|0\rangle\cdot \nonumber \\ &&
(2\pi)^2\delta (s_1)\delta (s_2) + \mbox{higher state contributions} 
\end{eqnarray}
In our approximation of massless quarks we put $m_{\pi}^2 = 0$.  So we see, that pion contribution into spectral
density is given by 
$\rho_{\mu\alpha\beta}^{\rm pion}\sim f_{\pi}^2 F_{\pi}(Q^2)p^{\alpha}p'^{\beta}(p^{\mu}+p'^{\mu})$
and as was already noted in \cite{Ioffe:ia,Ioffe:qb,Nesterenko:1982gc,Nesterenko:1984tk} the most convenient way to
extract pion form factor from QCD sum rules is to consider scalar amplitude in front of most symmetric
Lorentz structure $P_{\mu}P_{\alpha}P_{\beta}$ ($P=p+p'$).

\begin{figure}[ht]
\begin{center}
\includegraphics[scale=0.3]{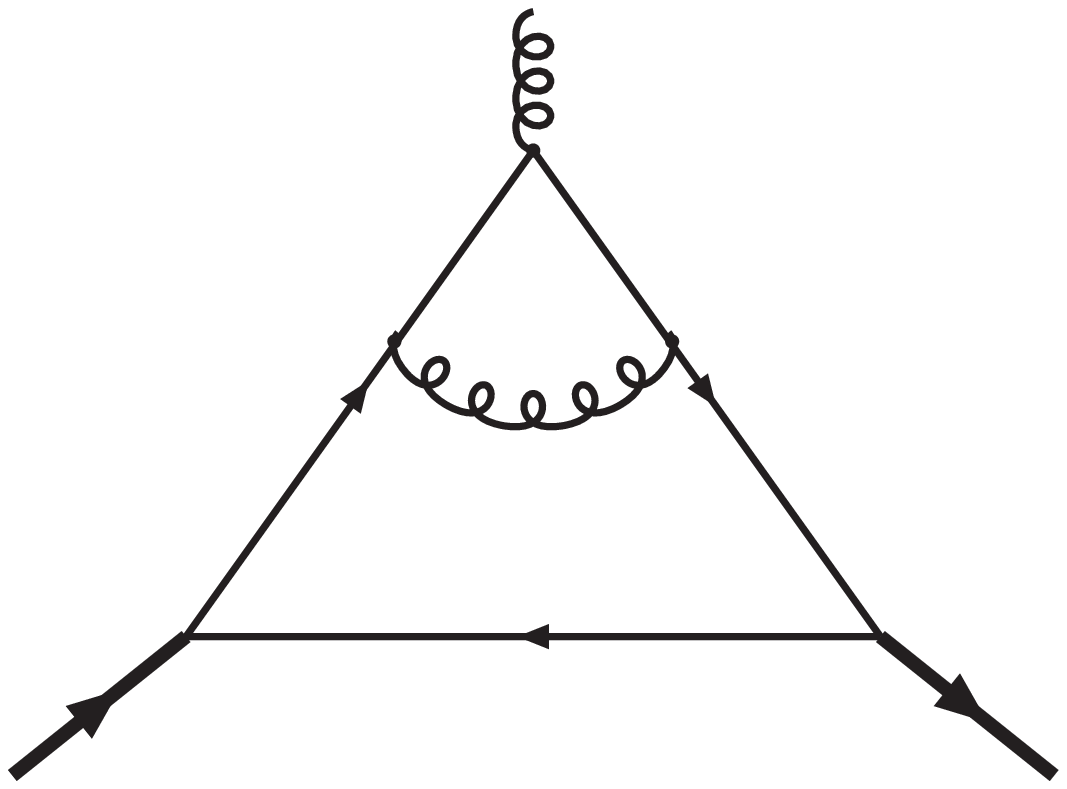} \makebox[2.cm]{}
\includegraphics[scale=0.3]{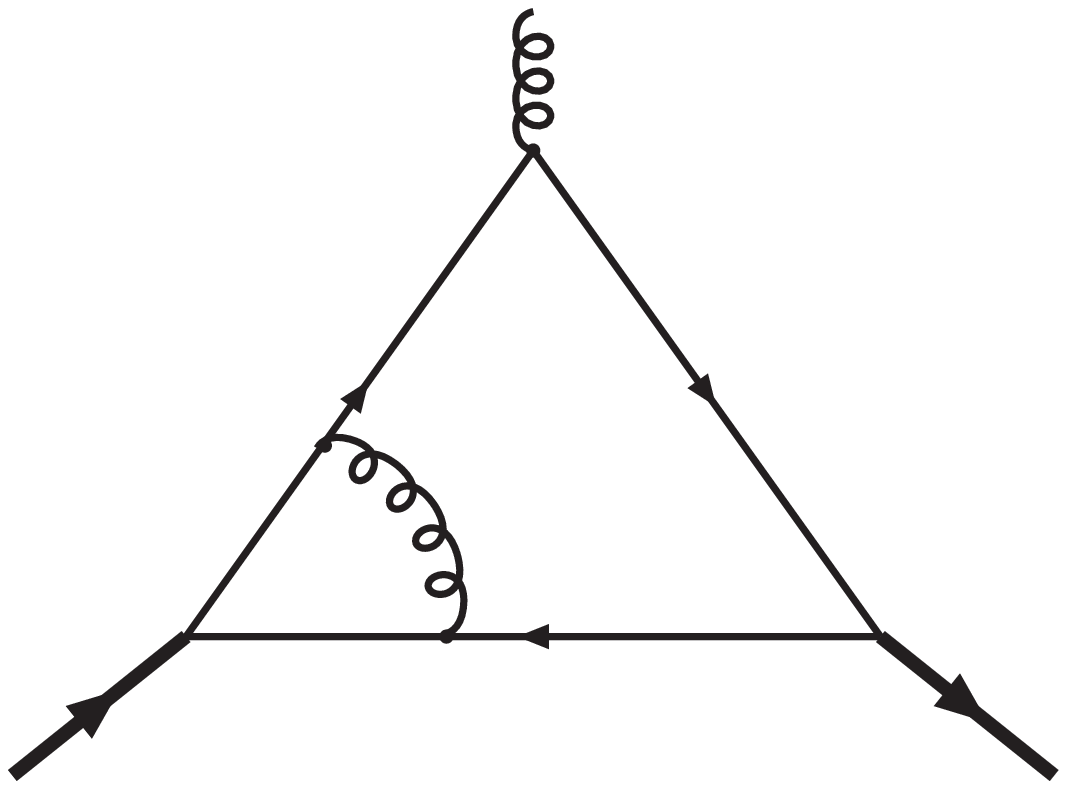} \makebox[2.cm]{} 
\includegraphics[scale=0.3]{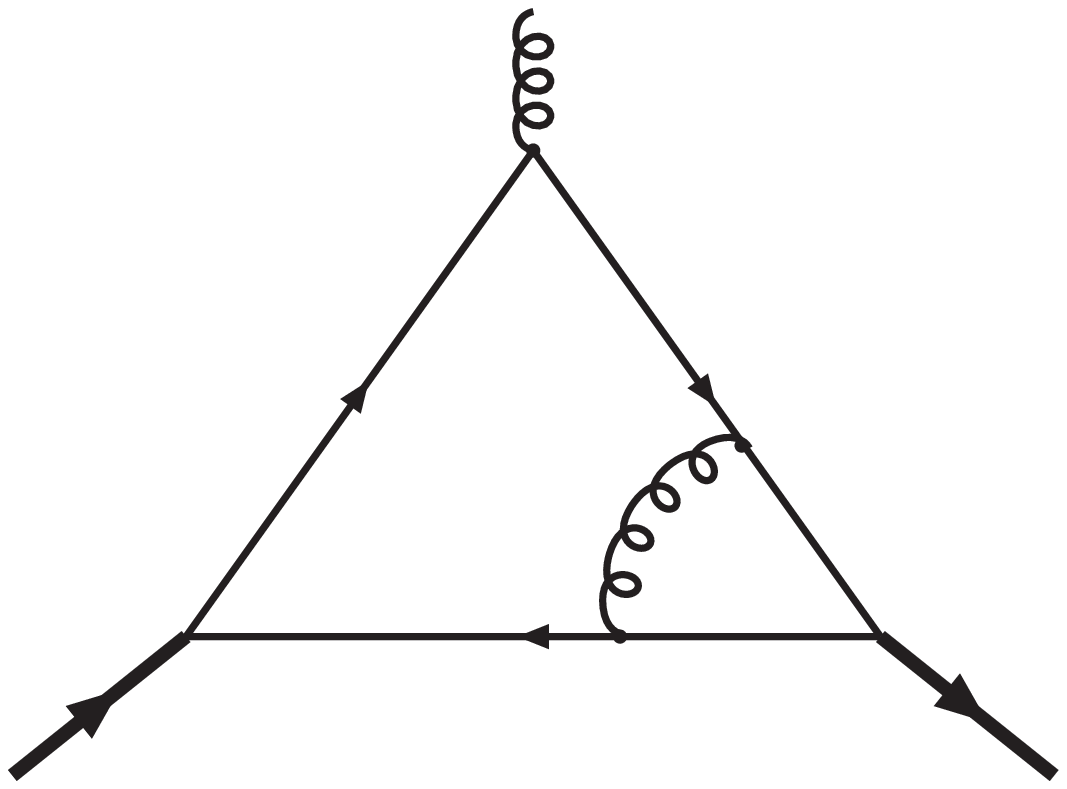} \\ \vspace*{1cm}
\includegraphics[scale=0.3]{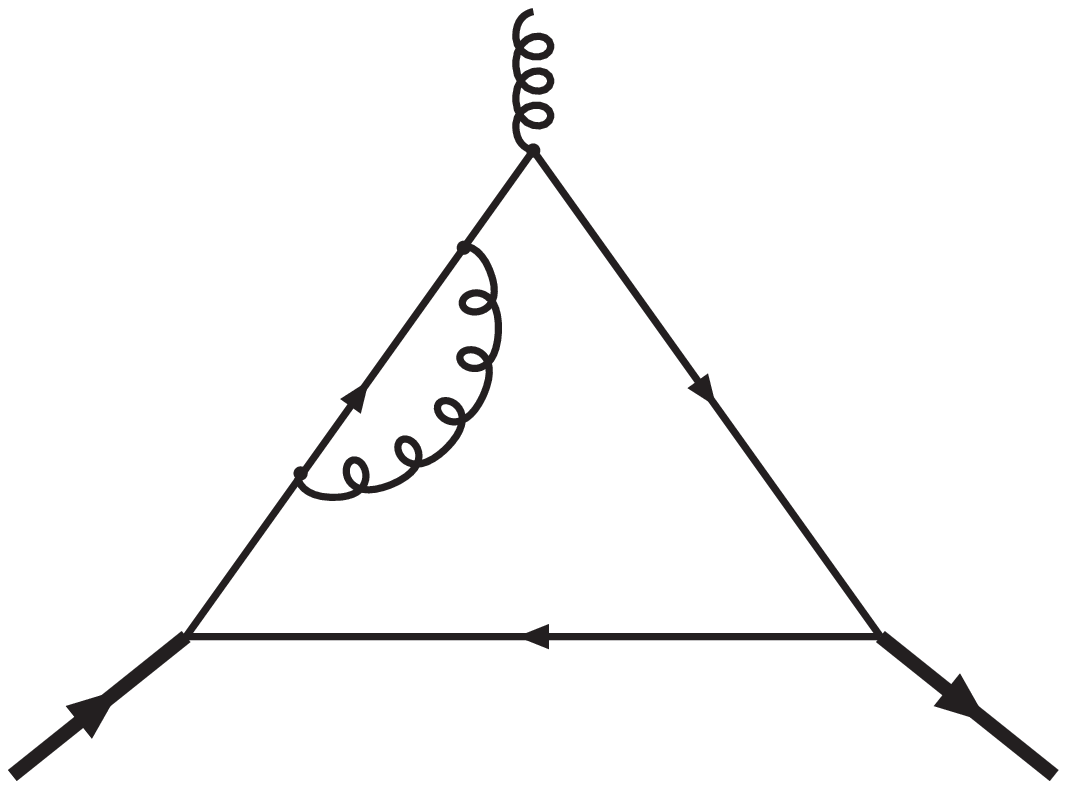} \makebox[2.cm]{}
\includegraphics[scale=0.3]{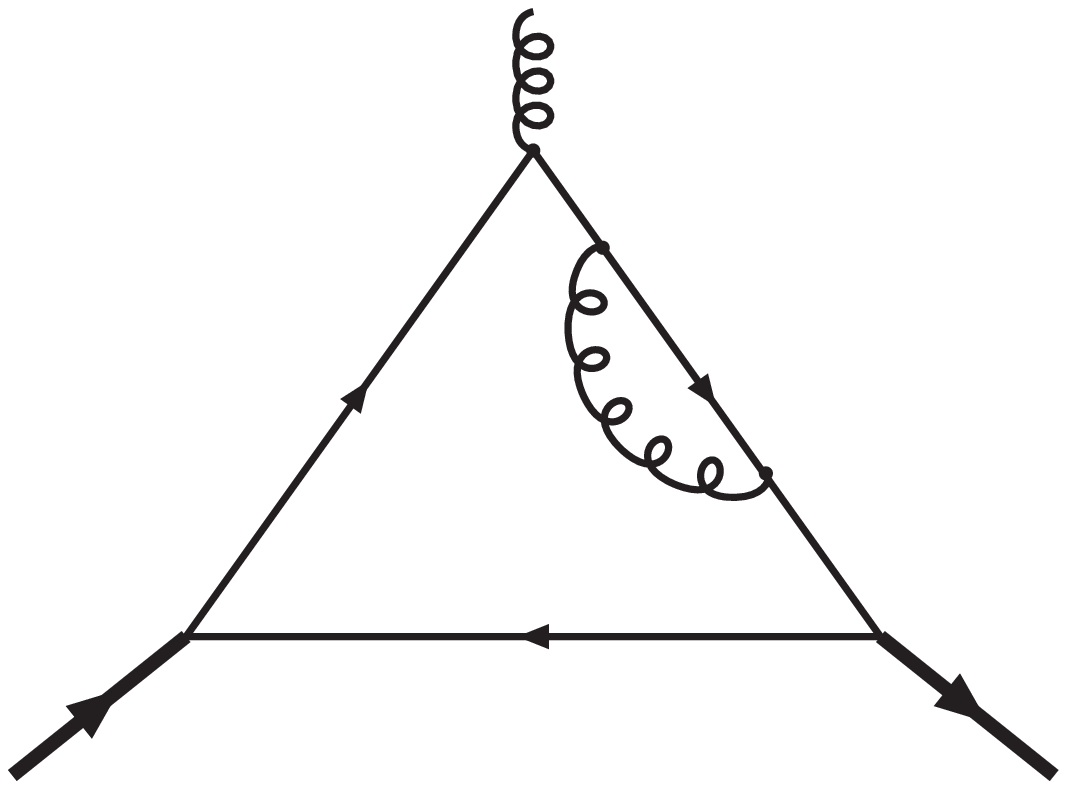} \makebox[2.cm]{} 
\includegraphics[scale=0.3]{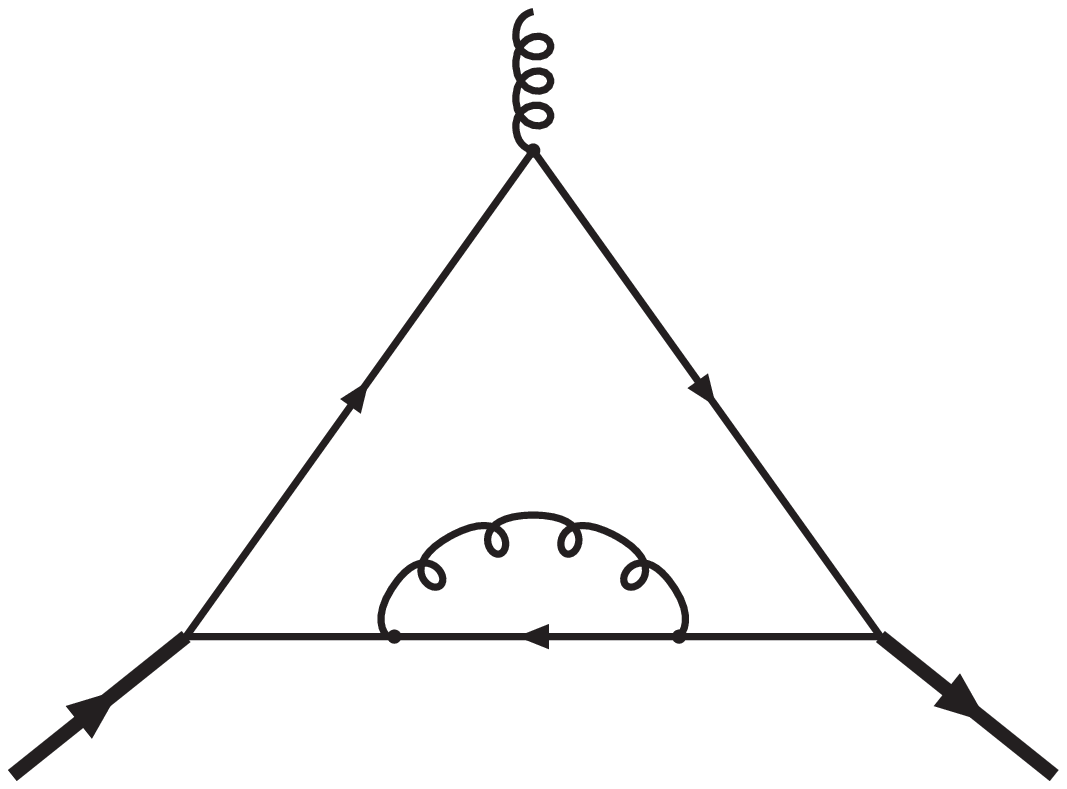}
\vspace*{0.5cm}
\caption{NLO diagrams}
\label{NLOdiagrams}
\end{center}
\end{figure}

Now it is time to return to the problem of evaluation of QCD expression for three point correlation function, 
we are interested in. The condensate contribution is known already for a long time 
\cite{Ioffe:ia,Ioffe:qb,Nesterenko:1982gc,Nesterenko:1984tk} and its analytical expression could
be found in the section with numerical results. The calculation of perturbative contribution could be 
conveniently performed with the use of double dispersion representation in variables $s_1 = p^2$ and $s_2 = p'^2$
at $q^2 < 0$:
\begin{eqnarray}
\Pi_{\mu\alpha\beta}^{\mathbf{pert}}(p^2, p'^2, q^2) = \frac{1}{(2\pi)^2}
\int\frac{\rho_{\mu\alpha\beta}^{\mathbf{pert}} (s_1,s_2,Q^2)}{(s_1-p^2)(s_2-p'^2)} ds_1ds_2 + \mbox{subtractions}
\label{doubledisp} 
\end{eqnarray} 
The integration region in (\ref{doubledisp}) is determined by condition
\footnote{In our case this inequality is satisfied identically.}
\begin{eqnarray}
-1 \leq \frac{s_2 - s_1 -q^2}{\lambda^{1/2} (s_1, s_2, q^2)} \leq 1
\end{eqnarray}
and
\begin{eqnarray}
\lambda (x_1, x_2, x_3) = (x_1 + x_2 -x_3)^2 - 4 x_1 x_2.
\end{eqnarray}
The double spectral density $\rho_{\mu\alpha\beta}^{\mathbf{pert}}(s_1,s_2,Q^2)$ is searched in the form of expansion
in strong coupling constant:
\begin{eqnarray}
\rho_{\mu\alpha\beta}^{\mathbf{pert}}(s_1,s_2,Q^2) = 
\rho_{\mu\alpha\beta}^{(0)}(s_1,s_2,Q^2) + \left(\frac{\alpha_s}{4\pi}\right)\rho_{\mu\alpha\beta}^{(1)}(s_1,s_2,Q^2)
+ \ldots
\end{eqnarray}

At leading order in coupling constant we have only one diagram depicted in Fig. 1, contributing to
three-point correlation function. At next to leading order we have 6 diagrams shown in Fig. 2.
The calculation of corresponding double spectral density was performed with the standard use
of Cutkosky rules. In the kinematic region $q^2 < 0$, we are interested in, there is no 
problem in applying Cutkosky rules for determination of $\rho_{\mu\alpha\beta} (s_1, s_2, Q^2)$ and integration
limits in $s_1$ and $s_2$. The non-Landau type singularities, not accounted for by Cutkosky
prescription, do not show up here. 

It is easy to find, that at Born level the scalar spectral density in front of most symmetric
Lorentz structure $P_{\mu}P_{\alpha}P_{\beta}$ is given by:

\begin{eqnarray}
\rho_{\mu\alpha\beta}^{(0)}(s_1, s_2, Q^2) &=& \frac {3 Q^4} 4 \frac 1 {k^{7/2}} \biggl(
3 k (s_1+s_2+Q^2)(s_1+s_2+2 Q^2)- k^2  \biggr. \nonumber \\
\biggl. && ~~~~~~~~~~~~~~~~~~~~~~~~~~~~~~ - 5 Q^2 (s_1+s_2+Q^2)^3 \biggr )P_{\mu}P_{\alpha}P_{\beta} + \ldots ,
\end{eqnarray}
\noindent
where $k = \lambda (s_1, s_2, -Q^2)$. The full analytical expression for $\rho_{\mu\alpha\beta}^{(0)}$
could be found in \cite{Ioffe:qb}. The calculation of NLO radiative corrections to double
spectral density is in principle straightforward. One just needs to consider all possible
double cuts of diagrams, shown in Fig. 2. However, the presence of collinear and soft infrared
divergences calls for appropriate regularization of arising divergences 
at intermediate steps of calculation and makes the whole analytical calculation quite
involved. We will present the details of NLO calculation in one of our future publications. Here
we give only final results. The calculation could be considerably simplified with the help of
Lorentz decomposition of double spectral density based on a fact, that our spectral density
is subject to three transversality conditions: 
$\rho_{\mu\alpha\beta}q_{\mu} = \rho_{\mu\alpha\beta}p_{\alpha} = \rho_{\mu\alpha\beta}p'_{\beta} = 0$: 
\begin{eqnarray}
\rho^{\mu\alpha\beta} &=&  A_1 [(Q^2+x)p_1^{\alpha}-(x+y)p_2^{\alpha}]
[(y-x)p_1^{\beta}+(Q^2+x)p_2^{\beta}][(Q^2+y)p_1^{\mu}+(Q^2-y)p_2^{\mu}] \nonumber \\
&& - \frac{1}{2}A_2 [(Q^2+y)p_1^{\mu}+(Q^2-y)p_2^{\mu}][(Q^2+x)g^{\alpha\beta}-2p_1^{\beta}p_2^{\alpha}] \nonumber \\
&& - \frac{1}{2}A_3 [(Q^2+x)p_1^{\alpha}-(x+y)p_2^{\alpha}][2(p_2^{\beta}-p_1^{\beta})p_2^{\mu}+(Q^2+y)g^{\mu\beta}]
\nonumber \\
&& - \frac{1}{2}A_4 [(x-y)p_1^{\beta}-(Q^2+x)p_2^{\beta}][2(p_2^{\alpha}-p_1^{\alpha})p_1^{\mu}+(y-Q^2)g^{\mu\alpha}],
\end{eqnarray}
where $x=s_1+s_2$ and $y=s_1-s_2$. The four independent structures 
$A_i$ (we suppressed the dependence on kinematical invariants) are given by a solution of system 
of linear equations: 
\begin{eqnarray}                
I_1 &=& \rho_{\mu \alpha \beta} p_1^{\mu} p_2^{\alpha} p_1^{\beta} = 
\frac {k^2} 8 \biggl ( k A_1 - A_2 - A_3 - A_4 \biggr )
\\
I_2 &=& \rho_{\mu \alpha \beta} p_1^{\mu} g^{\alpha \beta} = 
\frac{k}{4} (x+Q^2) \biggl ( k A_1 - 3 A_2 - A_3 - A_4 \biggr )
\\ 
I_3 &=& \rho_{\mu \alpha \beta} p_2^{\alpha} g^{\mu \beta} = 
\frac{k}{4} (y+Q^2) \biggl ( k A_1 - A_2 - 3 A_3 - A_4 \biggr )
\\ 
I_4 &=& \rho_{\mu \alpha \beta} p_1^{\beta} g^{\mu \alpha} = 
- \frac {k}{4} (y-Q^2) \biggl ( k A_1 - A_2 - A_3 - 3 A_4 \biggr ),
\label{system}
\end{eqnarray}         
The analytical expressions for $I_i$ (functional dependence on kinematical invariants is assumed) 
were found to be ($s_3 = Q^2$):
\begin{eqnarray}
k^{1/2}I_1 &=& -s_1^3 + s_2s_1^2 + s_2^2s_1-s_2^3 + (s_1+s_2)s_3^2 - s_1^2s_3 - s_2^2s_3 + s_1s_2s_3\Bigl[ \Bigr.\nonumber \\ 
&& - 16\log^2(v_1) - 16\log (v_3)\log (v_1) -16\log (v_4)\log (v_1)  \nonumber \\
&& + 2\log (v_1) - 4\log^2 (v_3) - 4\log^2 (v_4) - 2\log (v_2) - 2\log (v_3)
- 8\log (v_3)\log (v_4) \nonumber \\ && \left.
-8\Li_2\left(\frac{x_2}{x_1}\right) - 8\Li_2\left(\frac{y_1}{y_2}\right)
-8\Li_2\left(\frac{z_1}{s_1}\right) - 8\Li_2\left(\frac{z_1}{s_2}\right)
+8\Li_2\left(\frac{z_1}{z_2} \right)
\right], \\
k^{1/2}I_2 &=& - 2s_1^2 - 2s_2^2 + 2s_3^2 - 8s_1s_2 + s_1s_2\Bigl[ \Bigr. \nonumber \\
&& -32\log^2(v_1) - 32\log (v_3)\log (v_1) - 32\log (v_4)\log (v_1) + 4\log (v_1)  \nonumber \\
&&- 8\log^2 (v_3) - 8\log^2 (v_4) - 4\log (v_2) - 4\log (v_3) - 16\log (v_3)(v_4) \nonumber \\
&&\left. -16\Li_2\left(\frac{x_2}{x_1}\right) - 16\Li_2\left(\frac{y_1}{y_2}\right)
-16\Li_2\left(\frac{z_1}{s_1}\right) - 16\Li_2\left(\frac{z_1}{s_2}\right)
+16\Li_2\left(\frac{z_1}{z_2}\right)
\right], \\
k^{1/2}I_3 &=& -2s_1^2 + 2s_2^2 + 2s_3^2 -8s_2s_3 +s_2s_3\Bigl[ \Bigr. \nonumber \\
&& -32\log^2(v_1)-32\log (v_3)\log (v_1) -32\log (v_4)\log (v_1) + 4\log (v_1) \nonumber \\
&& -8\log^2(v_3) - 8\log^2(v_4) - 4\log (v_2) - 4\log (v_3) - 16\log (v_3)\log (v_4) \nonumber \\
&& \left.
-16\Li_2\left(\frac{x_2}{x_1}\right) - 16\Li_2\left(\frac{y_1}{y_2}\right)
-16\Li_2\left(\frac{z_1}{s_1}\right) - 16\Li_2\left(\frac{z_1}{s_2}\right)
+16\Li_2\left(\frac{z_1}{z_2}\right)
\right], \\
k^{1/2}I_4 &=& 2s_1^2 - 2s_2^2 + 2s_3^2 - 8s_1s_2 + s_1s_3\Bigl[ \Bigr. \nonumber \\
&& -32\log^2 (v_1) - 32\log (v_3)\log (v_1) - 32\log (v_4)\log (v_1) + 4\log (v_1) \nonumber \\
&& -8\log^2 (v_3) - 8\log^2 (v_4) - 4\log (v_2) - 4\log (v_3) - 16\log (v_3)\log (v_4) \nonumber \\
&& \left. 
-16\Li_2\left(\frac{x_2}{x_1}\right) - 16\Li_2\left(\frac{y_1}{y_2}\right)
-16\Li_2\left(\frac{z_1}{s_1}\right) - 16\Li_2\left(\frac{z_1}{s_2}\right)
+16\Li_2\left(\frac{z_1}{z_2}\right)
\right],
\end{eqnarray}
where the following notation was introduced:
\begin{eqnarray}
x_1 &=& \frac{1}{2}(s_1-s_2-Q^2)-\frac{1}{2}\sqrt{k}, \\
x_2 &=& \frac{1}{2}(s_1-s_2-Q^2)+\frac{1}{2}\sqrt{k}, \\
y_1 &=& \frac{1}{2}(s_1+Q^2-s_2)-\frac{1}{2}\sqrt{k}, \\
y_2 &=& \frac{1}{2}(s_1+Q^2-s_2)+\frac{1}{2}\sqrt{k}, \\
z_1 &=& \frac{1}{2}(s_1+s_2+Q^2)-\frac{1}{2}\sqrt{k}, \\
z_2 &=& \frac{1}{2}(s_1+s_2+Q^2)+\frac{1}{2}\sqrt{k}, \\
v_1 &=& \frac{1}{2s_1}(s_1-s_2-Q^2)+\frac{1}{2s_1}\sqrt{k}, \\
v_2 &=& \frac{1}{2s_2}(s_1-s_2+Q^2)+\frac{1}{2s_2}\sqrt{k}, \\
v_3 &=& \frac{1}{2s_1}(s_1+s_2+Q^2)+\frac{1}{2s_1}\sqrt{k}, \\
v_4 &=& \frac{s_1}{Q^2}, \\
v_5 &=& \frac{s_2}{Q^2}, \\
v_6 &=& 1 - \frac{z_1}{z_2}.
\end{eqnarray}
\noindent
We checked, that  all infrared and ultraviolet divergences cancel as should be for axial 
interpolating currents. Finally, NLO scalar spectral density in front of most symmetric Lorentz
structure $P_{\mu}P_{\alpha}P_{\beta}$ is  
\begin{eqnarray}
\rho_{\mu\alpha\beta}^{(1)} &=& \frac{Q^2}{k^{3}}\left\{
\frac{1}{2}(x_1+x_2)k I_3 + (k-5(x_1+x_2)(y_1+y_2)) I_1 \right. \nonumber \\
&& \left. 
+ \frac{1}{2}(y_1+y_2)k I_4 + \frac{1}{2}\frac{k}{z_1+z_2}((x_1+x_2)(y_1+y_2)-k)I_2 
\right\}P_{\mu}P_{\alpha}P_{\beta} + \ldots
\end{eqnarray}
In the limit $Q^2\to\infty$ our NLO double spectral density takes the following 
form:
\begin{eqnarray}
\rho_{\mu\alpha\beta}^{(1)} = 
\left\{
\frac{2}{Q^2} - 10\frac{s_1+s_2}{Q^4} - 2\frac{s_1+s_2}{Q^4}\log \left(\frac{s_1s_2}{Q^4} \right)
\right\}P_{\mu}P_{\alpha}P_{\beta} + \ldots
\end{eqnarray}
Here we would like to make several comments. First, we see that 
double logarithms are absent in our answer. Typically we would 
expect them to be nonzero as diagrams in Fig.\ref{NLOdiagrams} 
contain Sudakov vertex - corrections to $q$-vertex. And indeed our 
result for gluon correction to electromagnetic vertex (accompanied 
by the appropriate 1/2 self-energy insertions to ensure 
ultraviolet-finite result) contains double logs and agrees with large 
$Q^2$ limit of \cite{Bakulev1,Bakulev2}. However, our results for 
corrections to $p_1$ and $p_2$ vertexes also contain double logs 
(an analogue of double logs arising in pQCD description of pion 
electromagnetic form factor \cite{Botts:kf,Li:1992nu}). In the sum 
of all diagrams double logs cancel. Second, using large $Q^2$ 
expression for our spectral density and continuum subtraction thresholds for $s_1$ and $s_2$
equal to $4\pi^2f_{\pi}^2$  it is easy to see that leading $Q^2\to\infty$
QCD sum rule expression for pion electromagnetic form factor is given 
by leading pQCD prediction (\ref{pQCDLO}) with asymptotic pion distribution function 
($\phi_{\pi}^{as} (x) = 6f_{\pi}x (1-x)$) . Third, we checked numerically, that a $Q^2\to 0$
limit of our spectral density do satisfy Ward identity constrains
\footnote{We are grateful to A.P.Bakulev for explaining to us the details
of both this limit and Ward identity constrains for double spectral density} \cite{Bakulev3}.
Now let us proceed with numerical analysis.

\section{Numerical analysis}

In numerical analysis we used Borel scheme of QCD sum rules. That is, 
to get rid of unknown subtraction terms in (\ref{doubledisp}) we perform
Borel transformation procedure in two variables $s_1$ and $s_2$. The Borel
transform of three-point function $\Pi_i (s_1, s_2, q^2)$ is defined as
\begin{eqnarray}
\Phi (M_1^2, M_2^2, q^2)&\equiv & \hat B_{12}\Pi_i (s_1, s_2, q^2) =  \nonumber \\ &&
\lim_{n,m\to\infty}\left\{\left. \frac{s_2^{n+1}}{n!}
\left(-\frac{d}{d s_2}\right)^n \frac{s_1^{m+1}}{m!}
\left(-\frac{d}{d s_1}\right)\right|_{s_1=m M_1^2, s_2 = n M_2^2} \right\}
\Pi (s_1, s_2, q^2) \nonumber \\  \label{boreltransform}
\end{eqnarray}
Then Borel transformation (\ref{boreltransform}) of (\ref{doubledisp}) and (\ref{disp_phys}) gives 
\begin{eqnarray}
\Phi^{(\mathbf{pert}|\mathbf{phys})}  (M_1^2, M_2^2, q^2) = \frac{1}{(2\pi)^2}\int_0^{\infty}ds_1\int_0^{\infty}ds_2
\exp\left[-\frac{s_1}{M_1^2}-\frac{s_2}{M_2^2}\right]\rho^{(\mathbf{pert}|\mathbf{phys})} (s_1, s_2, q^2), 
\label{borelcor}
\end{eqnarray}
where $\rho^{(\mathbf{pert}|\mathbf{phys})} (s_1,s_2,q^2)$ stands for the expression of scalar spectral density in front 
of $P_{\mu}P_{\alpha}P_{\beta}$ Lorentz structure. In what follows we put $M_1^2 = M_2^2 = M^2$. 
If $M^2$ is chosen to be of order 1 GeV$^2$, 
then the right hand side of (\ref{borelcor}) in the case of physical spectral density will
be dominated by the lowest hadronic state contribution, while the higher state contribution 
will be suppressed.

\begin{figure}[ht]
\vspace*{3cm}
~~~~~~~~~~~~$F_{\pi}(1~\mbox{GeV}^2)$
\vspace*{-3cm}
\begin{center}
\includegraphics[scale=1.]{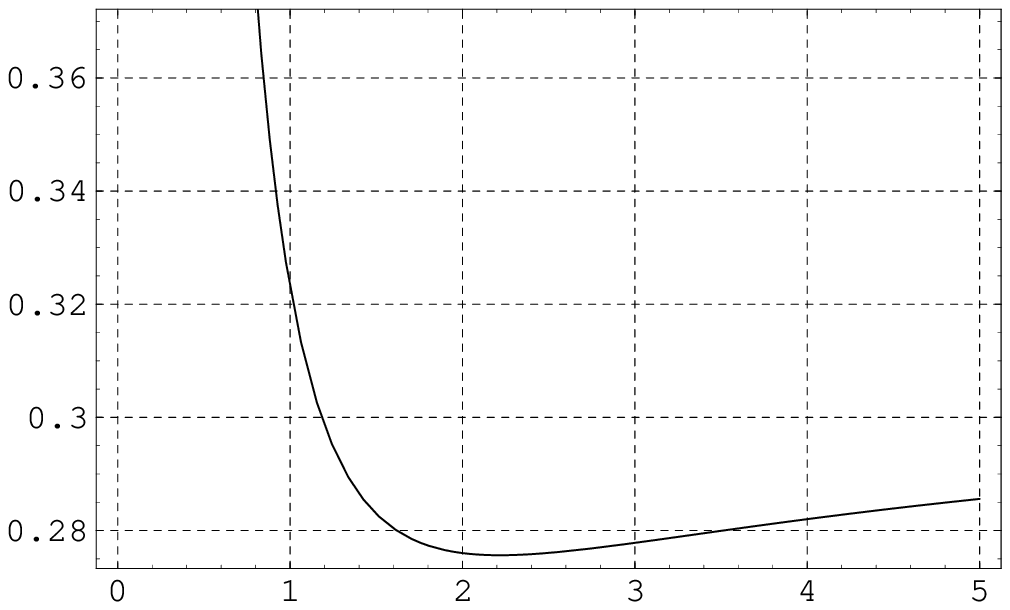} 
\caption{Borel mass $M^2$ dependence of pion electromagnetic form factor at $Q^2 =1~\mbox{GeV}^2$}
\label{formfactorplotM2}
\end{center}
\vspace*{-1.5cm}
~~~~~~~~~~~~~~~~~~~~~~~~~~~~~~~~~~~~~~~~~~~~~~~~~~~~~~~~~~~~~~~~~~~~~$M^2$
\vspace*{1.cm}
\end{figure}

Equating Borel transformed theoretical and physical parts of QCD sum rules we get
\begin{eqnarray}
F_{\pi} (Q^2) = \frac 4 { f_{\pi}^2 } \biggl ( \Phi (M^2, q^2) + \frac {\alpha_s} 
{48 \pi M^2} \langle 0| G_{\mu \nu}^a  G_{\mu \nu}^a |0 \rangle  +
\frac {52 \pi} {81 M^4 } \alpha_s 
\langle 0| \bar \psi \psi |0 \rangle^2 (1+\frac {2 Q^2} {13 M^2}) \biggr ), \label{formfactoreq}
\end{eqnarray}
where
\begin{eqnarray}
\Phi (M^2,q^2) = \frac{1}{(2\pi)^2}\int_0^{s_0} dx \exp\left[-\frac{x}{M^2}\right]
\int_0^x dy \rho^{\mathbf{pert}} (s_1,s_2,q^2).
\end{eqnarray}

\begin{figure}[ht]
\vspace*{3cm}
~~~~~~~~~~~~~~~~$Q^2F_{\pi}(Q^2)$
\vspace*{-3cm}
\begin{center}
\includegraphics[scale=1.]{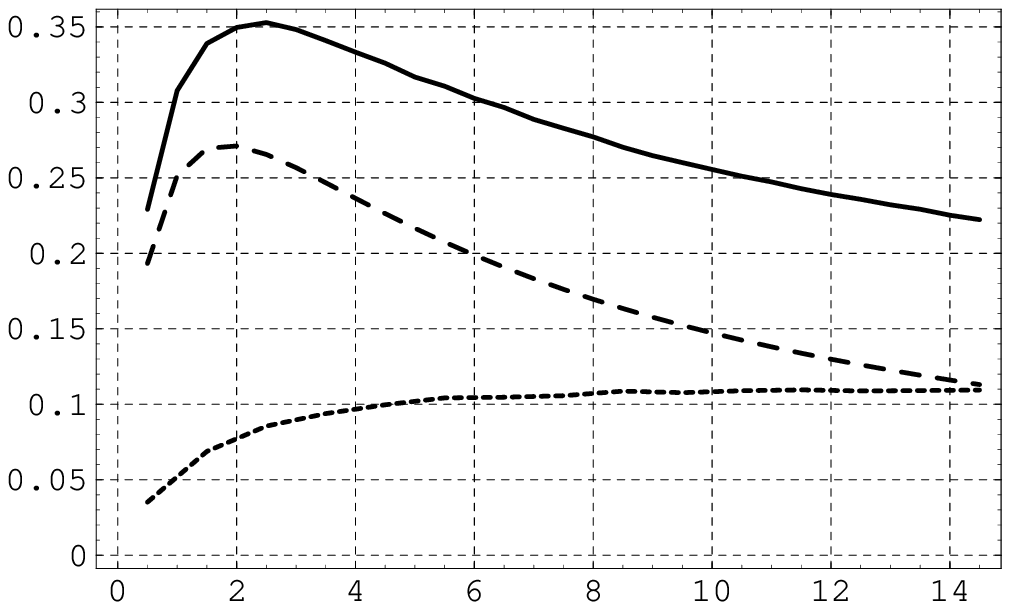} 
\caption{$Q^2$ dependence of pion electromagnetic form factor}
\label{formfactorplot}
\end{center}
\vspace*{-1.5cm}
~~~~~~~~~~~~~~~~~~~~~~~~~~~~~~~~~~~~~~~~~~~~~~~~~~~~~~~~~~~~~~~~~~~~~$Q^2$
\vspace*{1.cm}
\end{figure}

Here, for continuum subtraction we used so called "triangle" model. To verify the stability of our results with 
respect to the choice of continuum model we checked, that the usual "square" model gives similar predictions for pion 
electromagnetic form factor provided $s_0\sim 1.5 s_1$ is chosen\footnote{For more information about different continuum
subtraction models see \cite{Ioffe:qb}}. In what follows we use $s_0 = 0.96~\mbox{GeV}^2$ for continuum threshold
\footnote{In general, the value of continuum threshold is determined from the ratio of nonperturbative
contribution to leading perturbative term in OPE for our correlation function.}. 
This value is in agreement with the continuum threshold $\approx 0.7~\mbox{GeV}^2$ for axial polarization operator
used in two-point sum rules.  Next, we use two-loop renormalization group running of strong coupling constant 
with $\Lambda_{\rm QCD} = 325~\mbox{MeV}$ and fix the scale $\mu$ of strong coupling constant at $2~\mbox{GeV}$. This
choice is in agreement with the discussion presented in \cite{Bakulev2}, where it was argued that in the region of
momentum transfers $Q^2 < 10~\mbox{GeV}^2$ the strong coupling constant should be taken at frozen value $\alpha_s\sim 0.3$.   
In Fig. \ref{formfactorplotM2} we plotted the dependence of the pion electromagnetic form 
factor from the value of Borel parameter $M^2$ at $Q^2=1~\mbox{GeV}^2$. It is seen that "stability plateau" starts 
to develop for $M^2 > 2~\mbox{GeV}^2$. To proceed further we could fix the value of Borel parameter at 
$M^2 = 2~\mbox{GeV}^2$ and get results for pion form factor as a function of momentum transfer. However, doing so
we limit the range of $Q^2$ where our results could be considered as reliable. To see it, it is instructive to 
investigate large $Q^2$ limit of pion form factor. In the limit $Q^2\to\infty$ perturbative contribution to
pion form factor decreases as $Q^{-2}$, while power corrections grow with $Q^2$. It turns out, that our sum rules
become inapplicable at relatively large momentum transfers $Q^2 > 4$. Therefore, to cover all experimentally accessible
range of $Q^2$ for pion electromagnetic form factor we will explore our sum rules in the limit of infinite 
Borel parameter $M^2$.  So, basically, here we are employing local duality approach. The local duality
approach implies the following relation for continuum threshold \cite{Bakulev2,Bakulev3}: 
$s_1=s_2=4\pi^2f_{\pi}^2/(1+\frac{\alpha_s}{\pi}) =
0.62~\mbox{GeV}^2$, which ensures the Ward identity for pion electromagnetic form factor up to NLO ($F_{\pi}(0)=1$). 
The results for pion electromagnetic form factor are shown in Fig. \ref{formfactorplot}
(solid line is the sum of LO and NLO contributions, curve with long dashes denotes LO contribution
and curve with short dashes stands for NLO contributions.). 

Now, let us recall that in the limit $Q^2\to\infty$ our QCD sum rule prediction coincides with LO
pQCD results with asymptotic expression for pion distribution function. We could improve our prediction
further with inclusion of NLO pQCD corrections.   

\begin{figure}[ht]
\vspace*{3cm}
~~~~~~~~~~~~~~~~$Q^2F_{\pi}(Q^2)$
\vspace*{-3cm}
\begin{center}
\includegraphics[scale=1.]{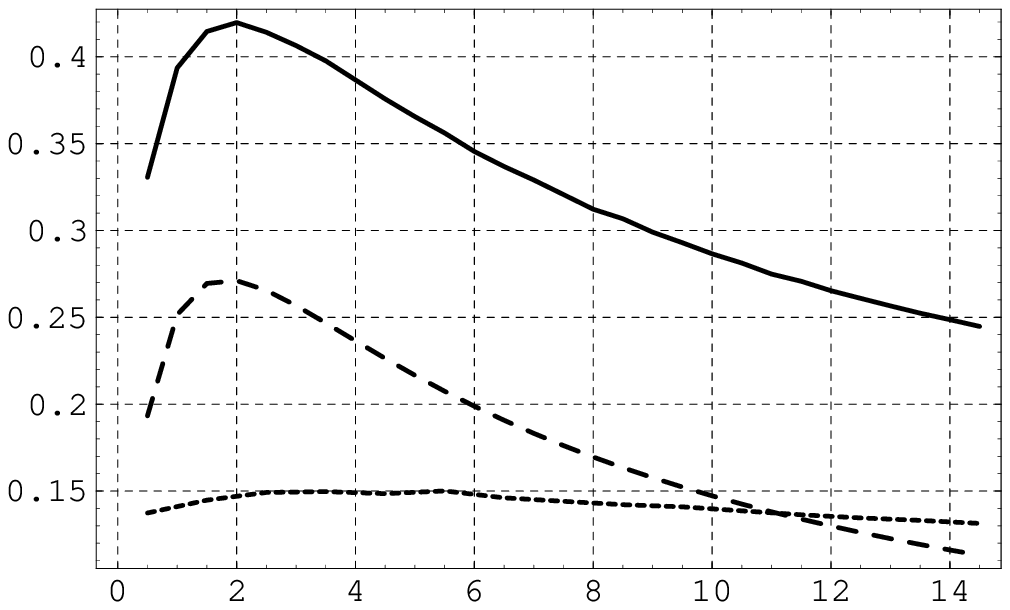} 
\caption{$Q^2$ dependence of pion electromagnetic form factor including NLO pQCD corrections}
\label{formfactorplotpQCD}
\end{center}
\vspace*{-1.5cm}
~~~~~~~~~~~~~~~~~~~~~~~~~~~~~~~~~~~~~~~~~~~~~~~~~~~~~~~~~~~~~~~~~~~~~$Q^2$
\vspace*{1.cm}
\end{figure}

Within perturbative QCD pion electromagnetic form factor is given by the following factorized expression
\begin{eqnarray}
F_{\pi}^{\mathbf{pQCD}} = \int_0^1 dx \int_0^1 dy \phi_{\pi} (x,\mu) T_H(x,y,Q^2,\mu^2)\phi_{\pi}(y,\mu),
\end{eqnarray}
with
\begin{eqnarray}
T_H (x,y,Q^2,\mu^2) = \frac{2\pi C_F\alpha_s (\mu)f_{\pi}^2}{N_c Q^2 (1-x)(1-y)}
\left[
1 + \frac{\alpha_s (\mu)}{\pi} T_1(x,y,Q^2/\mu^2) + \ldots
\right]
\end{eqnarray}
It should be noted that this contribution contains explicit dependence on pion wave function and gives us
a possibility for its determination through the comparison between combined predictions of QCD sum rules and pQCD
frameworks with experimental data. We suppose to perform this analysis in future, while in present work we 
limit ourselves only to the inclusion of NLO pQCD prediction with the use of asymptotic pion distribution functions.  
Within this approximation pQCD predictions for pion electromagnetic form factor are given by 
\cite{Field:wx,Dittes:aw,Khalmuradov:1984ij,Braaten:yy,Melic:1998qr} :
\begin{eqnarray}
F_{\pi}^{\mathbf{pQCD}} (Q^2) = F^{\mathbf{LO}} (Q^2) + F^{\mathbf{NLO}} (Q^2),
\end{eqnarray}
where
\begin{eqnarray}
F^{\mathbf{LO}} (Q^2) &=& 8\pi\alpha_s (\mu^2)\frac{f_{\pi}^2}{Q^2}, \\
F^{\mathbf{NLO}} (Q^2) &=&  8\frac{f_{\pi}^2}{Q^2}\alpha_s^2 (\mu^2)
\left[
6.58 + \frac{9}{4}\log\left(\frac{\mu^2}{Q^2}\right)
\right].
\end{eqnarray}
The results for pion electromagnetic form factor including NLO pQCD corrections are shown in Fig. \ref{formfactorplot}
(solid line is the sum of QCD sum rule and NLO pQCD contributions, curve with long dashes denotes LO QCD sum rule 
contribution and curve with short dashes stands for the sum of NLO QCD sum rule and NLO pQCD contributions.). 
We see, that obtained results for pion electromagnetic form factor are in good agreement with available experimental
data as well as already available theoretical estimates performed with the same level of precision.

\section{Conclusion}

We presented the results for pion electromagnetic form factor in the framework of three-point 
NLO QCD sum rules with axial interpolating currents for pions. To improve our estimates we also
took into account NLO pQCD contributions with asymptotic pion distribution functions. 
The theoretical curve obtained for $Q^2$ dependence of pion form factor is in a good agreement with
existing experimental data. Here, we for the first time computed radiative corrections
to double spectral density entering three-point sum rules with axial currents. These corrections turned
out to be large and should be taken into account in rigorous analysis.

We would like to thank A. Bakulev for interesting discussions and critical comments.
The work  of V.B. was supported in part by Russian Foundation of Basic Research under grant 01-02-16585, 
Russian Education Ministry grant E02-31-96, CRDF grant MO-011-0 and Dynasty foundation.  
The work of A.O. was supported by the National Science Foundation under 
grant PHY-0244853 and by the US Department of Energy under grant DE-FG02-96ER41005.

\end{document}